\documentclass[aps,physrev,reprint,noeprint]{revtex4-2} 
\usepackage{bm, braket,amsmath,mathtools,comment}
\usepackage{graphicx,color}
\usepackage{hyperref}

\begin{document}
\title{Matrix-product-state approach for qubits-waveguide systems in real space}
\author{Shimpei Goto}
\email[]{shimpei.goto@phys.s.u-tokyo.ac.jp}
\affiliation{Institute for Liberal Arts, Institute of Science Tokyo, Ichikawa, Chiba 272--0827, Japan}
\altaffiliation[present affiliation:]{ Department of Physics, The University of Tokyo, Tokyo 113--0033, Japan}
\date{\today}
\begin{abstract}
    We present a matrix-product-state-based numerical approach for simulating systems composed of several qubits and a common one-dimensional waveguide.
    In the presented approach, the one-dimensional waveguide is modeled in real space. Thus, one can use the advantage of matrix-product states that are suited for simulating low-entangled one-dimensional systems.
    The price to pay is that the vacuum of the waveguide in this modeling becomes the Bogoliubov vacuum, and one has to consider a not-so-small local Hilbert space for bosonic degrees of freedom.
    To manage the large local Hilbert space, we adopt the recently proposed single-site schemes.
    We demonstrate the potential of the presented approach by simulating superradiant phenomena within the Hamiltonian dynamics.
\end{abstract}
\maketitle
\section{Introduction\label{sec:introduction}}
It is a crucial configuration in superconducting circuits where multiple qubits share a common one-dimensional waveguide~\cite{jerger_readout_2011,jerger_frequency_2012,schmitt_multiplexed_2014,bultink_active_2016,neill_blueprint_2018,heinsoo_rapid_2018,spring_fast_2024,axline_-demand_2018,chang_remote_2020}.
By the shared waveguide, one can multiplex signals to qubits~\cite{jerger_readout_2011,jerger_frequency_2012,schmitt_multiplexed_2014,heinsoo_rapid_2018,spring_fast_2024} or transfer quantum states between qubits \cite{axline_-demand_2018,chang_remote_2020,cirac_quantum_1997}.
The multiplexing techniques reduce the number of cables connected to quantum computers based on superconducting circuits.
The transfer of quantum states leads to the remote connection of quantum computers.
Simulating and analyzing such configurations would improve the functionality of quantum computers.

Since it is quite difficult to experimentally observe wavefunctions, a comparison to the results obtained by numerical simulations is necessary to verify that quantum devices work as expected.
For such numerical simulations, tensor network simulations are suitable because of their quasi-exactness and applicability to the dynamics.
Among tensor networks, matrix product states (MPSs)~\cite{schollwock_density-matrix_2011} are a popular choice for the simulations of quantum devices~\cite{peropadre_nonequilibrium_2013,sanchez-burillo_scattering_2014,pichler_photonic_2016,di_paolo_efficient_2021,arranz_regidor_cavitylike_2021,arranz_regidor_modeling_2021,richter_enhanced_2022,arranz_regidor_probing_2023,ryu_efficient_2023,ryu_matrix_2023,papaefstathiou_efficient_2024,cilluffo_multimode-cavity_2024}.

For a system consisting of one or multiple qubits and a waveguide, numerical simulations based on an MPS representation are frequently performed~\cite{peropadre_nonequilibrium_2013,sanchez-burillo_scattering_2014,pichler_photonic_2016,arranz_regidor_cavitylike_2021,arranz_regidor_modeling_2021,richter_enhanced_2022,arranz_regidor_probing_2023,ryu_efficient_2023,ryu_matrix_2023}.
In these simulations, the waveguide is often modeled in frequency space~\cite{peropadre_nonequilibrium_2013,ryu_efficient_2023,ryu_matrix_2023}.
In the frequency-space representation, one can consider an arbitrary dispersion relation for the waveguide.
The price to pay is the interaction between qubits and the waveguide modes: A qubit is coupled to all waveguide modes.
Such couplings are long-range interactions in MPS representations, which often introduce distant entanglement and increase numerical costs.
This price is not expensive when a single qubit is coupled to the waveguide because a unitary transformation on the waveguide modes can make the couplings nearest-neighbor on a chain~\cite{prior_efficient_2010,chin_exact_2010,ryu_matrix_2023}.
In an \(N_q\)-qubit system, however, a similar unitary transformation introduces \(N_q\)-range couplings on a chain~\cite{ryu_efficient_2023,papaefstathiou_efficient_2024}.
As the number of qubits increases, an MPS simulation becomes more difficult because of long-range couplings.

If a waveguide is modeled in real space, such long-range couplings are absent as long as qubits are locally coupled to the waveguide.
Some studies perform numerical simulations based on a real-space waveguide indeed~\cite{sanchez-burillo_scattering_2014,calajo_exciting_2019}.
In these studies, the waveguide is modeled as a coupled resonator, i.e., the bosonic tight-binding chain.
Consequently, its dispersion relation is the ordinary cosine band and different from the linear dispersion in the standard waveguide.

In this paper, we introduce a real-space waveguide model whose dispersion relation can be recognized as linear and perform numerical simulations of the dynamics of a system composed of several qubits and the real-space waveguide model.
The simulations include both counter-rotating terms and non-Markovian effects.
The introduced real-space waveguide model is based on the quantum discrete transmission line~\cite{peropadre_nonequilibrium_2013,blais_cavity_2004,garcia_ripoll_quantum_2022}.
The vacuum of the introduced model is the Bogoliubov vacuum, and its elementary excitation is the linear Bogoliubov mode.
To describe the Bogoliubov vacuum, a relatively large local Hilbert space for bosonic degrees of freedom is required.
Because of this requirement, a real-space MPS approach based on the discrete transmission line is considered inefficient in Ref.~\cite{peropadre_nonequilibrium_2013}.
Due to the large local Hilbert space, the two-site update scheme routinely adopted in MPS simulations should be avoided.
The recently proposed controlled-bond expansion (CBE) approach~\cite{gleis_controlled_2023,li_time-dependent_2024} offers a single-site update scheme for ground-state searching and time evolution in the same manner. 
This approach significantly helps us to treat the large local Hilbert space.
By combining these techniques, we numerically simulate the dynamics of a system composed of qubits up to four and 400 waveguide modes.
The inefficient real-space MPS approach has become practical with recently developed techniques.
We observe the enhancement of the qubit decay rate with the number of qubits.
We also confirm that the enhancement follows the scaling expected from the superradiant phenomena.
The presented numerical approach can simulate the superradiant phenomena within the Hamiltonian dynamics.

The rest of the paper is organized as follows: The real-space waveguide model with linear dispersion is introduced in Sec.~\ref{sec:method}.
The details of the MPS approach is given in Sec.~\ref{sec:mps}.
In Sec.~\ref{sec:vacuum}, we investigate the dependence of the vacuum quality on the parameters used in simulations such as the dimension of the local Hilbert space.
The results of the simulation with a single qubit are presented in Sec.~\ref{sec:emission}.
In Sec.~\ref{sec:superradiant}, the results of the multi-qubit dynamics are shown.
The summary is given in Sec.~\ref{sec:summary}.
\section{Waveguide model in real space\label{sec:method}}
\begin{figure*}
    \includegraphics[width=0.95\linewidth]{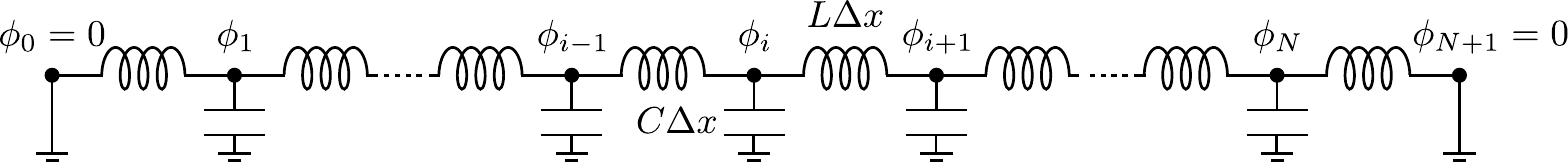}
    \caption{Circuit diagram for the discrete transmission line with the boundary condition \(\phi_0 = \phi_{N+1} = 0\).\label{fig:circuit}}
\end{figure*}

In the numerical approach presented in this paper, the Hamiltonian representation of a one-dimensional waveguide is obtained as the quantum discrete transmission line~\cite{blais_cavity_2004,garcia_ripoll_quantum_2022} with the boundary condition \(\hat{\phi}_0 = \hat{\phi}_{N+1} = 0\),
\begin{align}
    \label{eq:telegraph}
    \begin{aligned}
    \hat{H}_\mathrm{WG} = \ &\frac{1}{2C\Delta x}\sum^{N}_{i=1}\hat{q}^2_i + \frac{1}{2L\Delta x}\sum^{N-1}_{i=1}{(\hat{\phi}_{i+1} - \hat{\phi}_{i})}^2
    \\&+ \frac{1}{2L\Delta x}(\hat{\phi}^2_1 + \hat{\phi}^2_N),
    \end{aligned}
\end{align}
whose circuit diagram is shown in Fig.~\ref{fig:circuit}.
Here, \(N\) is the number of discretized points, \(C\) is the capacitance per length, \(L\) is the inductance per length, \(\Delta x\) is the discretized unit length, \(\hat{q}_i\) is the charge operator at the discretized point \(i\), and \(\hat{\phi}_i\) is the flux operator at the discretized point \(i\).
Since charge and flux are canonically conjugate quantities, the operators satisfies the following commutation relations~\cite{vool_introduction_2017}
\begin{align}
    \begin{aligned}
    [\hat{\phi}_i, \hat{q}_j] &= i\hbar \delta_{i,j}\\
    [\hat{q}_i, \hat{q}_j] &= 0\\
    [\hat{\phi}_i, \hat{\phi}_j] &= 0
    \end{aligned}
    .
\end{align}
Here, \(\delta_{i,j}\) is the Kronecker's delta.

The Hamiltonian representation of the waveguide~\eqref{eq:telegraph} can be recognized as coupled harmonic oscillators.
Thus, we define the bosonic annihilation operator for the local harmonic oscillator at the discretized point \(j\) as
\begin{align}
    \hat{a}_j = \sqrt{\frac{1}{2\hbar}\sqrt{\frac{2C}{L}}}\hat{\phi}_j + i \sqrt{\frac{1}{2\hbar}\sqrt{\frac{L}{2C}}}\hat{q}_j.
\end{align}
With the annihilation operator, the waveguide Hamiltonian~\eqref{eq:telegraph} is represented as
\begin{align}
    \label{eq:wg}
    \begin{aligned}
    \hat{H}_\mathrm{WG} = \sqrt{2}\hbar \omega \Bigg \{&\sum^N_{i=1}\left(\hat{a}^\dagger_i \hat{a}_i + \frac{1}{2}\right) \\
    &- \frac{1}{4}\sum^{N-1}_{i=1}(\hat{a}^\dagger_i + \hat{a}_i)(\hat{a}^\dagger_{i+1} + \hat{a}_{i+1}) \Bigg \} \\
    \end{aligned}
\end{align}
Here, \(\omega = 1/\sqrt{LC}\Delta x\).
Hereafter, we drop the constant terms in the Hamiltonian since they are irrelevant to dynamics. 
We use this representation for numerical simulations.

This Hamiltonian can be transformed into block diagonal form by representing in sine modes~\cite{garcia_ripoll_quantum_2022}
\begin{align}
    \hat{b}_n = \sqrt{\frac{2}{N+1}}\sum^N_{i=1} \sin  (k_n i \Delta x )\hat{a}_i 
\end{align}
with wavenumber \(k_n = \pi n / (N+1) \Delta x\).
With the sine modes, the annihilation operator in real space is decomposed as
\begin{align}
    \hat{a}_i = \sqrt{\frac{2}{N+1}}\sum^N_{n=1}\sin (k_n i \Delta x)\hat{b}_{n} 
\end{align}
and the Hamiltonian \(\hat{H}_\mathrm{WG}\) is represented as 
\begin{align}
    \begin{aligned}
    \hat{H}_\mathrm{WG} = \sqrt{2}\hbar \omega \sum^{N}_{n=1}\Bigg[&\left\{1 - \cos (k_n \Delta x) \right\}\hat{b}^\dagger_n \hat{b}_n \\&+ \frac{1}{4}\cos (k_n \Delta x)(\hat{b}^\dagger_n \hat{b}^\dagger_n + \hat{b}_n \hat{b}_n) \Bigg].
    \end{aligned}
\end{align}
Since anomalous terms like \(\hat{b}^\dagger_n \hat{b}^\dagger_n\) are present, we have to introduce the Bogoliubov modes
\begin{align}
    \hat{\alpha}_n = \hat{b}_n \cosh r_n  + \hat{b}^\dagger_n \sinh r_n
\end{align}
for the diagonalization.
By choosing the parameter \(r_n\) as
\begin{align}
    r_n = \frac{1}{4} \ln 2 + \frac{1}{2} \ln \sin \frac{k_n \Delta x}{2},
\end{align}
the Hamiltonian is transformed into its diagonal form 
\begin{align}
    \hat{H}_\mathrm{WG} = 2 \hbar \omega \sum^N_{n=1} \sin \frac{k_n \Delta x}{2} \hat{\alpha}^\dagger_n \hat{\alpha}_n.
    \label{eq:diagonal}
\end{align}
Consequently, the dispersion of the Bogoliubov mode is linear \(\hbar \omega k_n \Delta x\) when the wavenumber of the mode is small enough compared to the inverse of the discretized unit, i.e., \(k_n \Delta x \ll 1\).
As long as only energetically low-lying modes are excited, the Hamiltonian given in Eq.~\eqref{eq:wg} can be regarded as the real-space representation of one-dimensional waveguide whose speed of light \(c\) is given as \(c = \omega \Delta x = 1/\sqrt{LC}\).

With the obtained mode, straightforward calculations show that
\begin{align}
    \hat{a}^\dagger_j + \hat{a}_j = \frac{1}{2^{\frac{1}{4}}}\sqrt{\frac{2}{N+1}}\sum^N_{n=1} \frac{\sin(k_n j \Delta x)}{\sqrt{\sin\frac{k_n \Delta x}{2}}}(\hat{\alpha}^\dagger_n + \hat{\alpha}_n)\label{eq:inductive}
\end{align}
and
\begin{align}
    i(\hat{a}^\dagger_j - \hat{a}_j) &= 2^{\frac{1}{4}}i\sqrt{\frac{2}{N+1}} \nonumber\\ 
    &\times\sum^N_{n=1}\sin(k_n j \Delta x)\sqrt{\sin\frac{k_n \Delta x}{2}}(\hat{\alpha}^\dagger_n - \hat{\alpha}_n).
\end{align}
In short, the \(1/\sqrt{k_n}\) dependence of the inductive coupling and the \(\sqrt{k_n}\) dependence of the capacitive coupling~\cite{parra-rodriguez_quantum_2018} are automatically included in this approach.

\section{Matrix-product-states techniques\label{sec:mps}}
In the numerical simulations in this study, we represent a wavefunction \(\ket{\psi}\) in an MPS form~\cite{schollwock_density-matrix_2011}
\begin{align}
    \ket{\psi} = \sum_{\bm{\sigma}} \bm{A}^{\sigma_1}_1 \bm{A}^{\sigma_2}_2 \cdots \bm{A}^{\sigma_N}_N \ket{\bm{\sigma}},
\end{align}
where \(\sigma_i\) is the state of the local Hilbert space at the discretized point \(i\), \(\ket{\bm{\sigma}} = \bigotimes^{N}_{i=1}\ket{\sigma_i}\), and \(\sum_{\bm{\sigma}}\) means the summation over all possible configuration of \(\sigma_i\).
The matrix dimensions of matrices \(\bm{A}^{\sigma_i}_i\) are called bond dimensions.
The feasibility of numerical simulations based on the MPS representation is determined by the largeness of bond dimensions, or the entanglement of the wavefunction~\cite{schollwock_density-matrix_2011}.
This representation is efficient for low-entangled states such as the groundstates of spatially one-dimensional systems~\cite{calabrese_entanglement_2009}.
For the waveguide, we treat the bosonic degrees of freedom whose dimension of local Hilbert space is unbounded.
To perform numerical simulations, we choose local Fock states for the basis of the local Hilbert space and truncate an allowed occupation number down to a finite value \(N_{\mathrm{max}}\). 

As we will see in the next section, the highest occupation number \(N_{\mathrm{max}}\) should be \(O(10)\) to obtain a good quality vacuum state.
Because of the large local Hilbert space, standard MPS techniques~\cite{schollwock_density-matrix_2011,white_density_1992,white_density-matrix_1993,vidal_efficient_2003,vidal_efficient_2004,haegeman_time-dependent_2011,haegeman_unifying_2016} which update two sites at once are numerically expensive.
Thus, single-site update schemes~\cite{hubig_strictly_2015,yang_time-dependent_2020,gleis_controlled_2023,li_time-dependent_2024} are favorable for the simulations of systems considered in this study.
From such schemes, we adopt single-site density matrix renormalization group (DMRG) and time-dependent variable principle (TDVP) methods based on the CBE~\cite{gleis_controlled_2023,li_time-dependent_2024}.
During the CBE process, truncated singular value decomposition is performed twice: The first one is called pre-selection, and the second one is called final selection.
Throughout this study, we truncate singular values less than \(0.1\) (\(1.0 \times 10^{-3}\)) for pre-selection (final selection) in the expansion process.

To implement the DMRG and TDVP methods, the Hamiltonian \(H_\mathrm{WG}\) should be given in a matrix product operator (MPO) representation,
\begin{align}
    \label{eq:wg-MPO}
    \hat{H}_\mathrm{WG} = (0, 1, 0) \ \bm{W}_1 \bm{W}_2 \cdots \bm{W}_N \ \begin{pmatrix}
    1 \\ 0 \\ 0
    \end{pmatrix}.
\end{align}
Here, \(\bm{W}_i \) is a matrix whose coefficient is an operator acting on the discretized point \(i\).
By setting
\begin{align}
    \bm{W}_i =
    \begin{pmatrix}
        \hat{\openone}_i & 0 & 0 \\
        \sqrt{2}\hbar \omega \hat{a}^\dagger_i \hat{a}_i & \hat{\openone}_i & - \frac{\sqrt{2}}{4}\hbar \omega(\hat{a}^\dagger_i + \hat{a}_i) \\
        \hat{a}^\dagger_i + \hat{a}_i & 0 & 0
    \end{pmatrix},
\end{align}
one can obtain the waveguide Hamiltonian~\eqref{eq:wg} (except irrelevant constants).
Here, \(\hat{\openone}_i\) is the identity operator for the discretized point \(i\).

\section{Vacuum state of waveguide\label{sec:vacuum}}
Since the waveguide Hamiltonian~\eqref{eq:wg} contains anomalous terms, the vacuum for the annihilation operator \(\hat{a}_i\) is not the vacuum for the Bogoliubov modes \(\hat{\alpha}_n\).
Even for considering a situation where no excitation is present in the waveguide, one has to prepare an entangled state.
This is the drawback of simulating the waveguide model in real space.
One can construct the vacuum for the Bogoliubov modes from their amplitudes using the expression given in Ref.~\onlinecite{ma_multimode_1990}.
However, the states generated from this expression are slightly different from the eigenstates of the Hamiltonian used in numerical simulations because of the truncation of bosonic occupation number.

As inferred from the diagonal form of the waveguide Hamiltonian~\eqref{eq:diagonal}, the vacuum state for the Bogoliubov modes is obtained as the ground state of the waveguide Hamiltonian~\eqref{eq:wg}.
Since the waveguide Hamiltonian is described as the one-dimensional chain, its ground state would be efficiently described by a MPS.
By using the DMRG with CBE~\cite{gleis_controlled_2023}, we obtain the ground state of the waveguide Hamiltonian~\eqref{eq:wg} with some highest occupation number \(N_\mathrm{max}\) and number of discretized points \(N\).
In the DMRG calculations, we truncate singular values less than \(10^{-10}\) after the update of a single matrix \(\bm{A}^{\sigma_i}_i\).
To assess how the parameters \(N_\mathrm{max}\) and \(N\) affect the quality of a vacuum, we evaluate the occupation number of each Bogoliubov mode \(\braket{\hat{\alpha}^\dagger_n \hat{\alpha}_n}\) with the ground state obtained by the DMRG method.

\begin{figure}
    \includegraphics[width=0.75\linewidth]{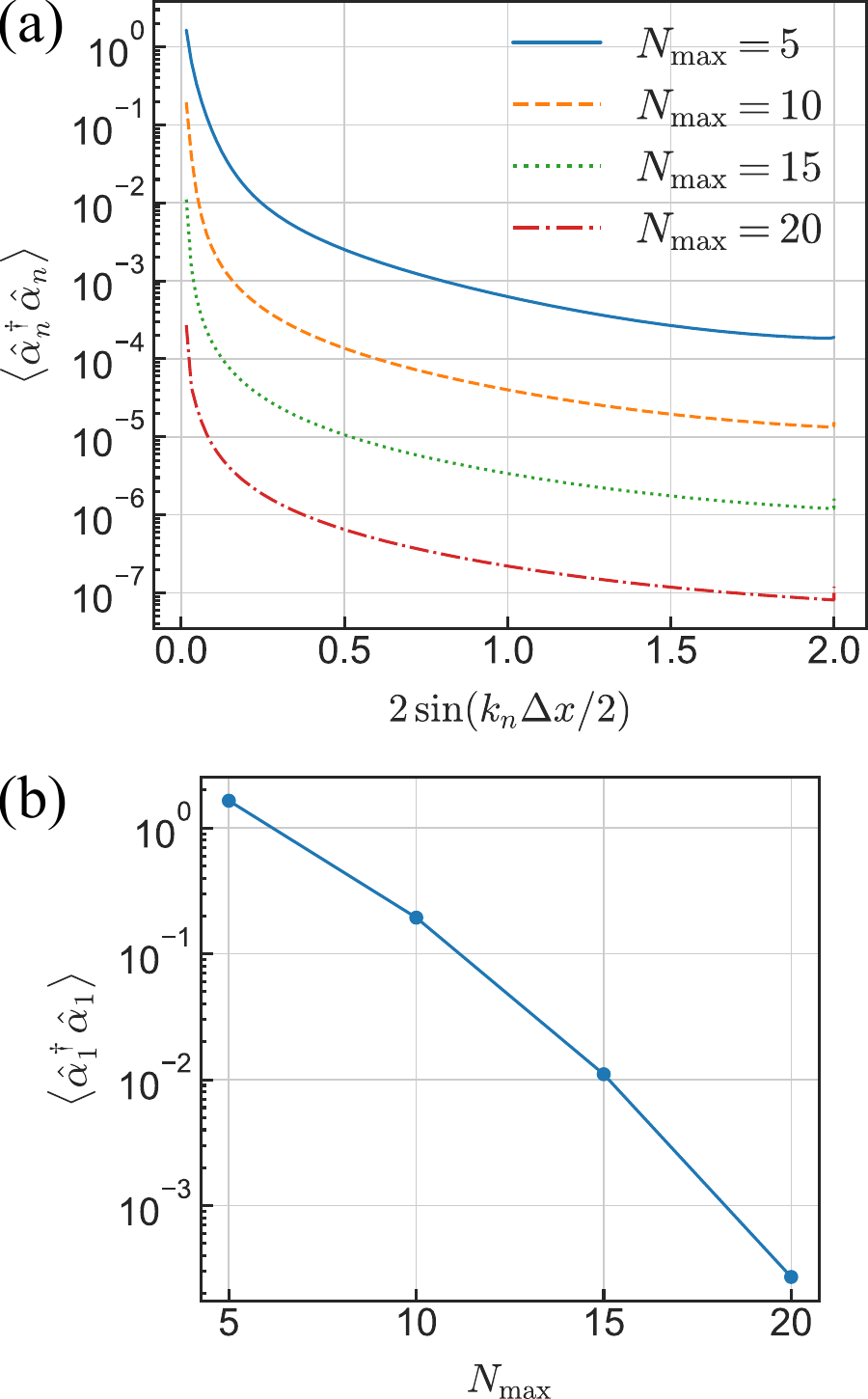}
    \caption{(a) Excitation energy dependence of the occupancy of each Bogoliubov mode for different highest occupation numbers \(N_\mathrm{max}\). We set the number of discretized points \(N\) to \(200\). (b) Highest occupation number dependence of the occupancy of the energetically lowest Bogoliubov mode \(\braket{\hat{\alpha}^\dagger_1 \hat{\alpha}_1}\). The line is a guide for the eye.\label{fig:nmax_dep}}
\end{figure}

We see the effects of the highest occupation number \(N_\mathrm{max}\) first. 
Figure~\ref{fig:nmax_dep}(a) shows the excitation energy dependence of the occupancy of each Bogoliubov mode for different highest occupation numbers \(N_\mathrm{max}\).
The excitations of Bogoliubov modes remain finite in the ground state of the Hamiltonian \(\hat{H}_\mathrm{WG}\) obtained by the DMRG method, and the occupancy becomes larger as the excitation energy decreases.
The remaining occupancy becomes smaller as the highest occupation number \(N_{\mathrm{max}}\) increases.
This fact supports that the finite excitations come from the truncation of the bosonic Hilbert space.
Figure~\ref{fig:nmax_dep}(b) gives the highest occupation number \(N_\mathrm{\max}\) dependence of the occupancy of the energetically lowest Bogoliubov mode \(\braket{\hat{\alpha}^\dagger_1 \hat{\alpha}_1}\).
We observe the super-exponential decay of the occupancy with the increase of the highest occupation number. 
The occupancies of energetically low-lying Bogoliubov modes can be suppressed by setting \(N_{\mathrm{max}}\) to around \(20\).

\begin{figure}
    \includegraphics[width=0.95\linewidth]{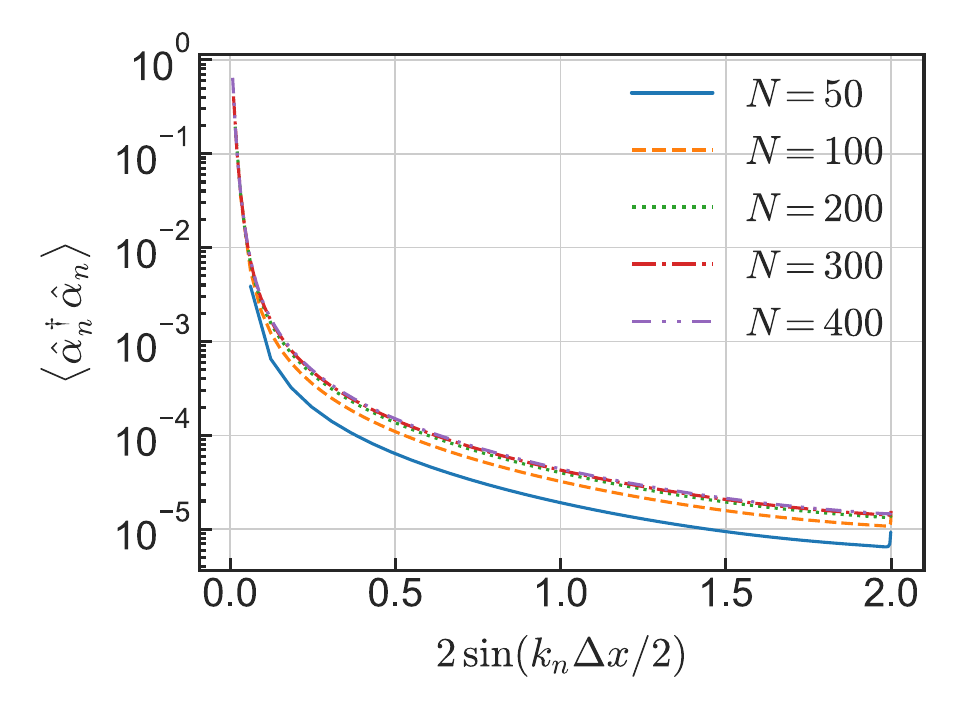}
    \caption{Excitation energy dependence of the occupancy of each Bogoliubov mode for different numbers of discretized points \(N\). We set the maximum occupation number \(N_{\mathrm{max}}\) to 10.\label{fig:n_dep}}
\end{figure}

Next, we see how the number of discretized points \(N\) affects the occupancy of the Bogoliubov modes.
Figure~\ref{fig:n_dep} represents the excitation energy dependence of the occupancy of each Bogoliubov mode for different numbers of discretized points \(N\).
For smaller number of discretized points \(N\), the occupancy becomes larger as \(N\) increases.
This increase, however, stops and the occupancy shows little dependence on the number of discretized points \(N\) for sufficiently large region \(N \geq 200\).
The excitation energy and the highest occupation number determine the occupancy of each Bogoliubov mode in such a region.
We note that the lowest excitation energy is determined by the number of discretized points \(N\).
The dependence of the parameter \(N\) appears thorough the lowest excitation energy.

From these observations, the quality of a vacuum state obtained by the DMRG method is essentially determined by the highest occupation number \(N_{\mathrm{occ}}\).
The remaining occupancy is larger for energetically lower-lying Bogoliubov modes.
The parameter \(N_\mathrm{max}\) should be chosen so that the occupancy of relevant Bogoliubov modes are sufficiently suppressed.

\section{Single qubit dynamics\label{sec:emission}}

With the waveguide model whose character is investigated in previous sections, we construct the Hamiltonian of qubits interacting with a common one-dimensional waveguide,
\begin{align}
    \label{eq:qubit-wg}
    \hat{H} = \sum^{N_q}_{n=1} \left\{\frac{\hbar \omega_{q, n}}{2} \hat{Z}_n + \hbar g_n \hat{X}_n(\hat{a}^\dagger_{i_n} + \hat{a}_{i_n})\right\} + \hat{H}_{\mathrm{WG}}.
\end{align}
Here, \(N_q\) is the number of qubits, \(\omega_{q, n}\) is the resonant frequency of the \(n\)-th qubit, \(\hat{Z}_n\) and \(\hat{X}_n\) are the Pauli-\(Z\) and \(X\) operators acting onto the \(n\)-th qubit, respectively, \(g_n\) denotes the inductive coupling between the \(n\)-th qubit and the waveguide, and \(i_n\) is the discretized point where the \(n\)-th qubit is coupled. 
Qubits are modeled as two-level systems.
One can obtain the MPO representation of this qubits-waveguide Hamiltonian~\eqref{eq:qubit-wg} from the MPO representation of the waveguide Hamiltonian~\eqref{eq:wg-MPO} by inserting a matrix \(\bm{Q}_{n}\) in front of the \(\bm{W}_{i_n}\),
where
\begin{align}
    \bm{Q}_n = 
    \begin{pmatrix}
        \hat{\openone}_n & 0 & 0 \\
        \frac{\hbar \omega_{q,n}}{2}\hat{Z}_n & \hat{\openone}_n & \hbar g_n \hat{X}_n \\
        0 & 0 & \hat{\openone}_n
    \end{pmatrix}.
\end{align}
By inserting the matrix for the qubit \(n\), the nearest neighbor couping between \(i_n -1\) and \(i_n\) becomes next nearest neighbor coupling.
This modification of interaction range does not depend on the number of qubits \(N_q\) unlike frequency-space approaches.
Hereafter, we consider cases where all the qubit frequencies \(\omega_{q,n}\) and the couplings \(g_n\) do not depend on the qubit index \(n\), i.e., \(\omega_{q, n} = \omega_q\) and \(g_{n} = g\).

With the qubits-waveguide Hamiltonian~\eqref{eq:qubit-wg}, we obtain the time-evolved state
\begin{align}
    \ket{\psi(t)} = \exp\left(-\frac{i}{\hbar}\hat{H}t\right)\ket{\psi(0)},
\end{align}
where \(\ket{\psi(0)}\) is an initial state.
The dynamics we try to simulate is the spontaneous emission from excited qubits.
Consequently, an initial state is given by
\begin{align}
    \ket{\psi(0)} = \prod^{Nq}_{n=1} \exp \left(-i\frac{\pi}{2}\hat{Y}_n \right) \ket{\psi_0},
\end{align}
where \(\hat{Y}_n\) is the Pauli-\(Y\) operator acting onto the \(n\)-th qubit and \(\ket{\psi_0}\) is the ground state of the qubit-waveguide Hamiltonian~\eqref{eq:qubit-wg} obtained by the DMRG method.
The quantity we are interested in is the population of \(n\)-th qubit
\begin{align}
    P_n (t) = \frac{\braket{\psi(t)|\hat{Z}_n |\psi(t)}+1}{2}.
\end{align}
In the Markovian dynamics, the population is expected to show exponential decay with decay rate \(\gamma_n\), \(P_n(t) \sim e^{-\gamma_n t}\).
Based on this expectation, we evaluate the (time-dependent) decay rate as
\begin{align}
    \gamma_n (t) = -\frac{1}{P_n(t)}\frac{dP_n(t)}{dt}.
\end{align}
The time derivative of the population is evaluated by using the expression
\begin{align}
    \frac{dP_n(t)}{dt} = g \braket{\psi(t)|\hat{Y}_n(\hat{a}^\dagger_{i_n} + \hat{a}_{i_n})|\psi(t)}.
\end{align}
This relation is obtained from the Heisenberg equation of motion.

We start our simulations from a simple single-qubit case where only a single qubit is coupled to a waveguide.
The couping \(g\) is set to \(7.0 \times 10^{-2} \omega_q\).
We set the waveguide frequency \(\omega \) to \(40 \omega_q / 2\pi \).
This value is equivalent to \(\Delta x = \textcolor{red}{\lambda_q} / 40\), i.e., the discrete unit length \(\Delta x\) is set to 1/40 of the wavelength of the excitation resonant to the qubit frequency \(\textcolor{red}{\lambda_q = 2\pi c / \omega_q}\).
The number of discretized point \(N\) is set to \(200\), and the qubit is coupled to the 100th point.
The dynamics is simulated by the TDVP method with CBE~\cite{li_time-dependent_2024}.
In the TDVP calculations, we truncate singular values less than \(10^{-8}\) after the update of a single matrix.
The time step for simulating the dynamics is set to \(0.025 \omega^{-1}_q\).

\begin{figure}
    \includegraphics[width=0.95\linewidth]{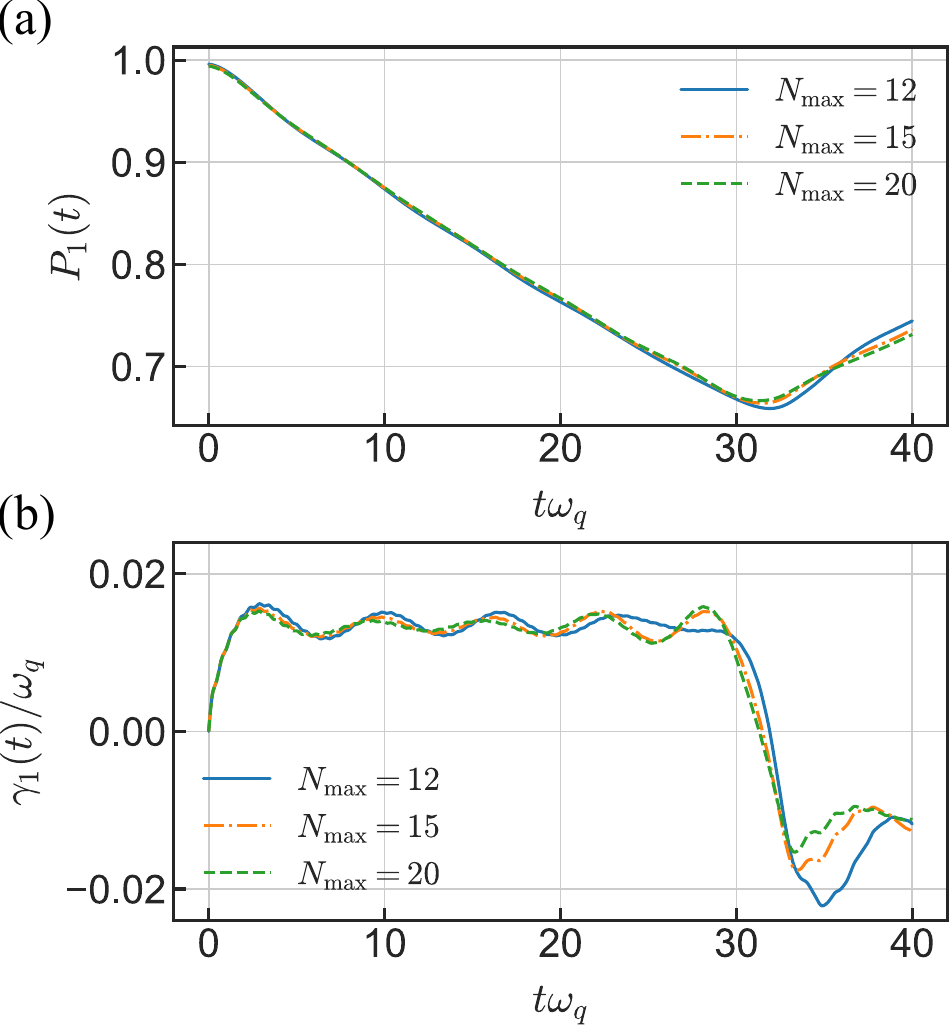}
    \caption{Time evolution of (a) the population and (b) the decay rate of the single-qubit coupled to the waveguide. The waveguide model is composed of 200 discretized points and the qubit is coupled to the 100th point. We use three highest occupation numbers 12, 15, and 20 in the simulations.\label{fig:single_qubit}}
\end{figure}
Figure~\ref{fig:single_qubit}(a) shows the time evolution of the qubit population \(P_1(t)\) for different highest occupation numbers 12, 15, and 20.
The populations with different highest occupation numbers do not show notable difference up to \(t \sim 30 / \omega_q \).
For the region \(t \gtrsim 30 / \omega_q\), the populations start to increase and one can observe notable difference between the case with \(N_\mathrm{max} = 12\) and the others.
This revival of the population is caused by the finiteness of the waveguide. 
Emitted excitation is reflected at the edge of waveguide and comes back to the qubit.
Since the qubit is attached to the center of the waveguide, the round-trip time for the emitted excitation \(t_r\) is estimated to be \(N \Delta x / c = N/\omega \).
This value is \(10 \pi \) with the parameters used in the simulations.
The estimated round-trip time is consistent with time points where the revivals of the populations starts.
The consistency also confirms that the velocity of excitations in the waveguide is \(c\) as designed.

The time evolution of the decay rate is shown in Fig.~\ref{fig:single_qubit}(b).
The decay rate shows an oscillating behavior in relatively early time and becomes positive in late time.
Such temporal changes of the decay rate is recognized as non-Markovian behaviors. 
By averaging the decay rate over the period \(10/\omega_q \leq t \leq 20 / \omega_q\), the decay rate for single qubit \(\gamma \) is estimated to be \(\gamma/\omega_q = 1.36 \times 10^{-2}\).
Hereafter, we use this value for the single qubit decay rate.
The difference between the case with \(N_\mathrm{max} = 12\) and the others is more visible compared to the population even for \(t \lesssim 30 / \omega_q\).
Despite the visible difference, the simulation with \(N_\mathrm{max} = 12\) is sufficient for estimating the time-averaged decay rate before the revival.
If one has interest in precise temporal behaviors of the decay rate, the highest occupation number should be set to around 20.
An appropriate value for \(N_\mathrm{max}\) depends on required preciseness.

\section{Multi-qubit dynamics\label{sec:superradiant}}

In this section, we simulate systems with several numbers of qubits.
Except for the number of discretized points, the other settings, including simulation parameters, are the same as the single-qubit case in the previous section.

\subsection{Correlation between two distant qubits}
To demonstrate the validity of our approach for describing non-Markovian effects, we simulate systems with two distant qubits whose detailed results are reported in Refs.~\onlinecite{alvarez-giron_delay-induced_2024,regidor_theory_2025}.
The number of discretized points is increased to \(1000\), and the first qubit is coupled to the 410th discretized point.
The second qubit is coupled to the 570th point or the 590th point.
When the second qubit is connected to the 570th (590th) point, the distance between the two qubits corresponds to \(4 \lambda_q\) (\(4.5 \lambda_q\)) and the propagating phase \(\phi = \omega_q l / c\) is \(8\pi \) (\(9 \pi \)).
Here, \(l\) is the distance between the two qubits.
The parameter \(\gamma l/c\) that characterizes the dynamics is approximately 0.3418 and 0.3845 for \(\phi = 8\pi \) and \(9\pi \), respectively.

\begin{figure}
    \includegraphics[width=0.98\linewidth]{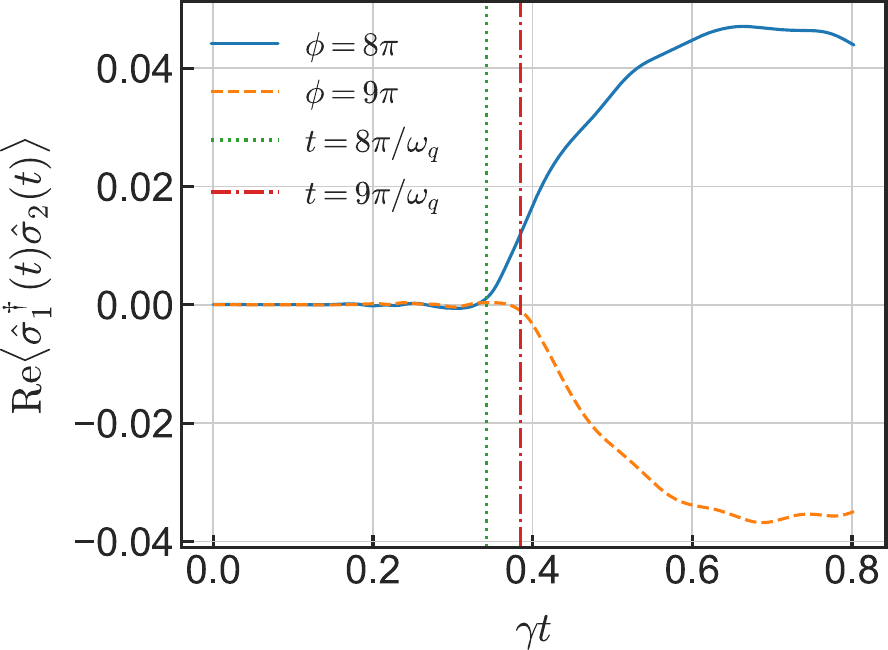}
    \caption{Time evolution of the correlation \(\braket{\hat{\sigma}^\dagger_1(t)\hat{\sigma}_2(t)}\) between two distant qubits. The vertical lines represent the duration required for propagating emitted photons between the qubits. The waveguide model is composed of 1000 discretized points. The first qubit is coupled to the 410 th point. The second qubit is connected to the 570th (590th) point for the propagating phase \(\phi = 8 \pi \) (\(9\pi \)).\label{fig:correlation}}
\end{figure}

The time evolution of the correlation \(\braket{\hat{\sigma}^\dagger_1 \hat{\sigma}_2}\) between the two qubits is presented in Fig.~\ref{fig:correlation}.
Here, the operator \(\hat{\sigma}_j\) is defined as \((\hat{X}_j - i \hat{Y}_j)/2 \).
The correlation remains almost zero until a photon emitted from one qubit reaches the other qubit (except for small fluctuations that might come from the truncation of the wavefunction).
The situation we simulate is slightly different from that performed in Ref.~\cite{regidor_theory_2025}: Our simulation includes counter-rotating terms~\footnote{The rotating-wave approximation introduces long-range interactions in the present real-space approach.} and the wavenumber dependence of the inductive coupling as shown in Eq.~\eqref{eq:inductive}.
Despite these differences, the correlation obtained from our simulation with \(\phi = 9\pi \) that corresponds to \(\gamma l / c = 0.3845\) shows good agreement with the correlation for \(\gamma l / c = 0.375\) shown in Fig.~4 of Ref.~\cite{regidor_theory_2025}.
Specifically, the peak value read from Fig.~4 of Ref.~\cite{regidor_theory_2025} is around -0.038, and that of our simulation is around -0.037.
This agreement assures the validity of our approach for describing non-Markovian dynamics.

Moreover, Ref.~\cite{regidor_theory_2025} has reported that the sign of the correlation depends on the propagating phase \(\phi = m \pi \).
The parity of integer \(m\) determines the sign of the correlation.
As shown in Fig.~\ref{fig:correlation}, our simulation reproduces this parity dependence of the correlation.
This fact also supports the validity of our approach.

\subsection{Enhancement of decay rates}
Next, we consider the cases where qubits are placed within a small region.
The number of discretized point is set to \(400\), and the \(n\)-th qubit is coupled to the \((199+n)\) th discretized point.

\begin{figure}
    \includegraphics[width=0.95\linewidth]{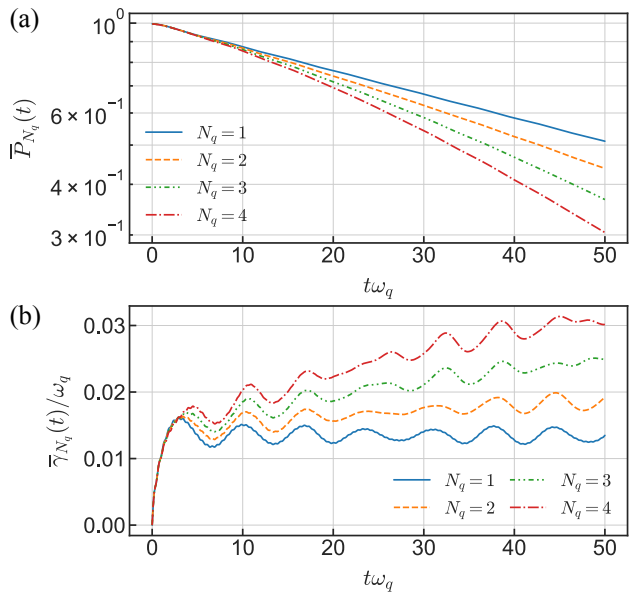}
    \caption{Time evolution of the averaged (a) populations and (b) decay rates of qubits coupled to the waveguide. We also plot the averaged decay rate of the Markovian dynamics with \(N_q = 2\) for comparison. The waveguide model is composed of 400 discretized points. The \(n\)-th qubit is coupled to the \((199+n)\)-th point. The highest occupation number is set to 12 in the simulations.\label{fig:multi_qubit}}
\end{figure}

Figure~\ref{fig:multi_qubit} represents the time evolution of the averaged qubit population \(\overline{P}_{N_q}(t)\) defined by
\begin{align}
\overline{P}_{N_q}(t) = \frac{1}{N_q}\sum^{N_q}_{n=1}P_n(t)
\end{align}
and the averaged decay rate \(\overline{\gamma}_{N_q}(t)\) given by
\begin{align}
\overline{\gamma}_{N_q}(t) = \frac{1}{N_q}\sum^{N_q}_{n=1}\gamma_n(t)
\end{align}
for several numbers of qubits \(N_q\).
For the Markovian dynamics with \(N_q = 2\), the simple expression for the qubit population
\begin{align}
    P_{n, \mathrm{Markov}}(t) = (1 + \gamma t) e^{-2 \gamma t}
\end{align}
is available~\cite{gross_superradiance_1982}.
From this expression, one can obtain the averaged decay rate for the Markovian dynamics as
\begin{align}
    \overline{\gamma}_{2, \mathrm{Markov}} = \gamma \left( 2 - \frac{1}{1 + \gamma t}\right).
\end{align}
The obtained averaged decay rate \(\overline{\gamma}_2(t)\) oscillates around the Markovian value \(\overline{\gamma}_{2, \mathrm{Markov}}(t)\) with the single qubit decay rate \(\gamma / \omega_q = 1.36 \times 10^{-2}\).
This oscillating behavior also assures the validity of our approach.

As the number of qubits \(N_q\) increases, the averaged population shows more rapid decrease.
The averaged decay rate also increases with the number of qubits.
The enhancement of the decay rate coming from the multi-qubit character can be recognized as the superradiant phenomena~\cite{dicke_coherence_1954}.

\begin{figure}
    \includegraphics[width=0.95\linewidth]{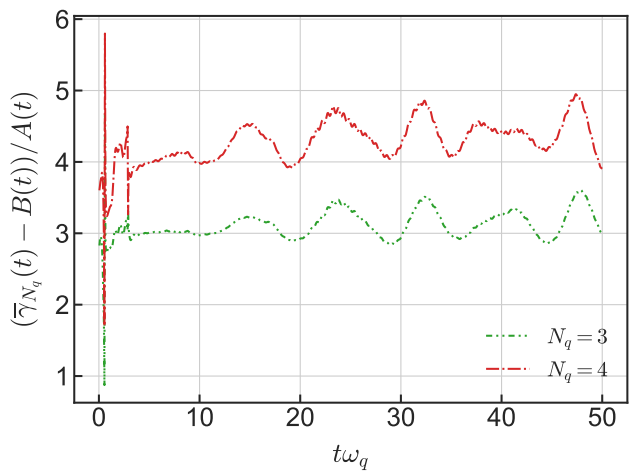}
    \caption{Rescaled and shifted averaged decay rate for \(N_q=3\) and \(4\). The time-dependent rescaling value \(A(t)\) and shift value \(B(t)\) are determined from the averaged decay rate for \(N_q=1\) and \(2\).\label{fig:Nq}}
\end{figure}

The superradiant phenomena is characterized by the quadratic scaling of the total photon emission rate with the number of qubits \(N_q\).
In terms of the averaged decay rate, the linear scaling with \(N_q\) is the characteristics of the supperradiant phenomena.
Since the obtained decay rate has time dependence coming from the non-Markovian effects, the expected scaling of the averaged decay rate with \(N_q\) is given by
\begin{align}
    \overline{\gamma}_{N_q}(t) \sim A(t)N_q + B(t).
    \label{eq:gamma_scaling}
\end{align}
Here, \(A(t)\) and \(B(t)\) are time-dependent coefficients.
To confirm the averaged decay rate follows the expected scaling, we estimate the time-dependent coefficients using \(\overline{\gamma}_1(t)\) and \(\overline{\gamma}_2(t)\) as
\begin{align}
    A(t) = \overline{\gamma}_2(t) - \overline{\gamma}_1(t)
\end{align}
and
\begin{align}
    B(t) = 2\overline{\gamma}_1(t) - \overline{\gamma}_2(t).
\end{align}
With the estimated coefficients, we obtain the rescaled and shifted averaged decay rate \((\overline{\gamma}_{N_q}(t) - B(t))/A(t)\) for \(N_q = 3\) and \(4\) as shown in Fig.~\ref{fig:Nq}.
Except for early time, the rescaled and shifted averaged decay rate behaves as \((\overline{\gamma}_{N_q}(t) - B(t))/A(t) \sim N_q\).
The averaged decay rate obtained by the numerical simulations follow the scaling expected by the superradiant phenomena \eqref{eq:gamma_scaling}.
In other words, the real-space approach can treat collective interaction between qubits properly.

\begin{figure}
    \includegraphics[width=0.95\linewidth]{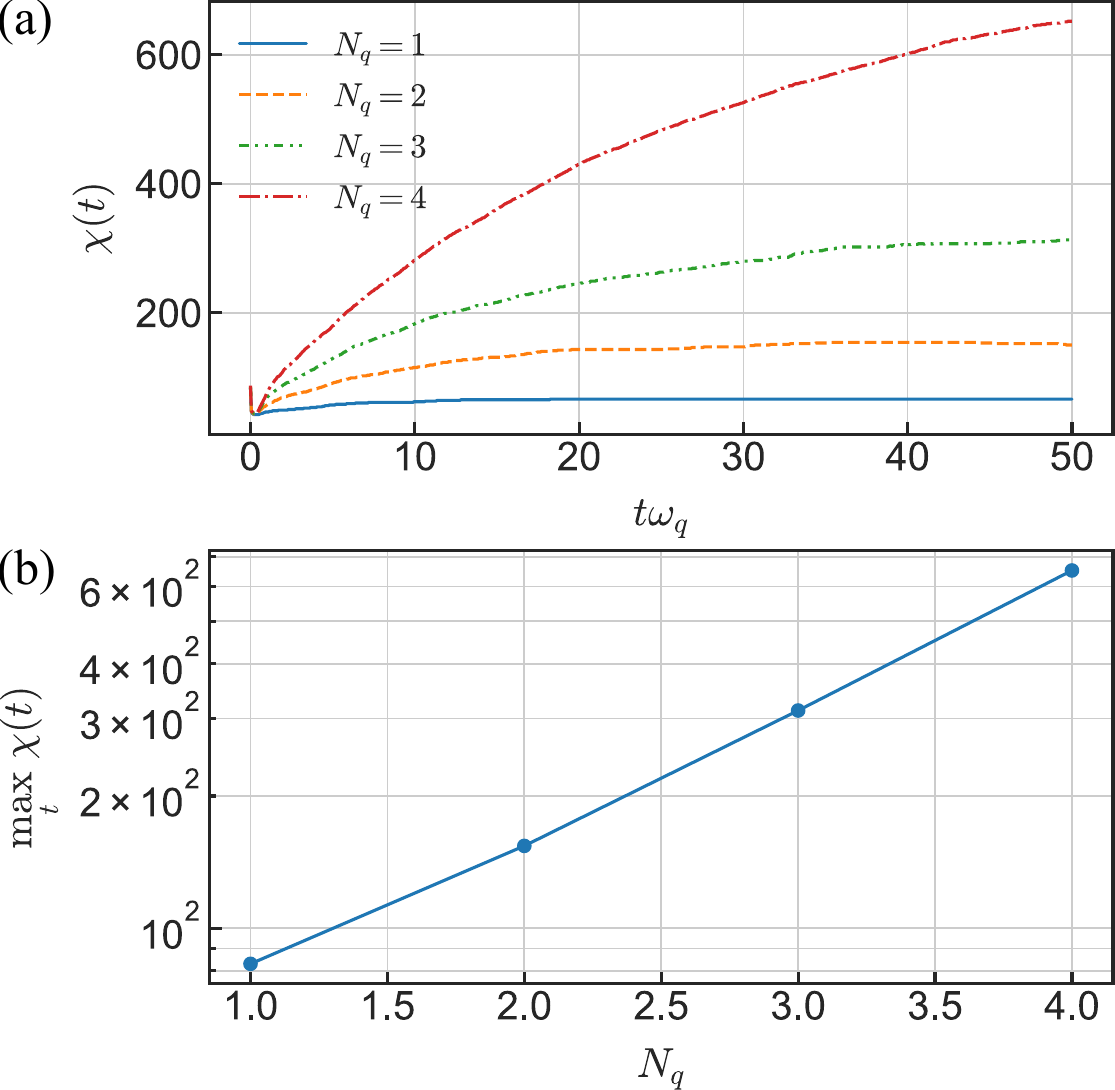}
    \caption{(a) Time evolution of the largest bond dimension \(\chi(t)\) in the MPS representation. (b) Number of qubits dependence of the maximized largest bond dimension \(\max_t \chi(t)\).\label{fig:bond_dim}}
\end{figure}

The numerical cost of MPS simulations can be inferred from the bond dimension, the matrix dimension of matrices \(\bm{A}^{\sigma_i}_i\).
When the largest bond dimension in the matrices \(\bm{A}^{\sigma_i}_i\) is \(\chi \), the biparted entanglement entropy can be bounded by \(\ln \chi \).
Figure~\ref{fig:bond_dim} represents the time evolution of the largest bond dimension and the number of qubits dependence of the maximal value.
The bond dimension increases sub-linearly with time.
As far as the emission dynamics simulated in this section, the evolution of entanglement is not so severe.
On the other hand, the maximized largest bond dimension shows exponential scaling with the number of qubits \(N_q\).
This exponential increase with \(N_q\) indicates that the volume-law-scaling entanglement is present among qubits.
Consequently, numerical cost increases polynomially with time but exponentially with number of qubits. 
The presented MPS-based approach is appropriate for simulating several qubits interacting via a common one-dimensional waveguide, as long as entanglement among qubits is small enough to be accessible by an MPS representation.

\section{Summary\label{sec:summary}}
Based on the quantum discrete transmission line description of a one-dimensional waveguide~\cite{blais_cavity_2004,garcia_ripoll_quantum_2022}, we performed the numerical simulations of the emission dynamics of qubits coupled to a common one-dimensional waveguide in real-space representation.
One of advantages of the real-space representation is the absence of long-range interaction whose range depends on the number of qubits.
The drawback is that the vacuum state for the waveguide is an entangled Bogoliubov vacuum.
Besides, many local Fock states are required to describe a good-quality vacuum.
We overcome this difficulty by using recently proposed controlled bond expansion approaches~\cite{gleis_controlled_2023,li_time-dependent_2024}.
With the presented method, we performed the numerical simulations of the emission dynamics of systems composed of several qubits and a common waveguide.
We confirmed that the presented method can reproduce the dynamics of the correlation between two remote qubits reported in Ref.~\cite{regidor_theory_2025}.
We also observed that the method can describe the collective phenomena between qubits, the superradiance.

To access analytical expressions presented in Sec.~\ref{sec:method}, we considered only the spatially homogeneous cases in this paper.
The waveguide model used in this paper can be easily extended to spatially inhomogeneous cases, such as periodic capacitance that corresponds to periodic transmittivity.
The discretized length can also be spatially dependent.
By setting the discretized length long near the boundaries, it would be possible to postpone the arrival of a reflected excitation.
The utilization of spatial dependence is an interesting direction for extending the presented approach.

In the simulations in this paper, the counter-rotating terms are included, and the entanglement between qubits and waveguide modes is taken into consideration.
Since these often-neglected contributions are present, the presented approach is well-suited for simulating superconducting circuits in the ultrastrong (or stronger) coupling regime~\cite{niemczyk_circuit_2010,forn-diaz_ultrastrong_2017,yoshihara_superconducting_2017,bosman_multi-mode_2017,kuzmin_superstrong_2019,ao_extremely_2023,tomonaga_spectral_2025}, where these contributions play key roles.
Compared to the frequency-space approach based on the discrete transmission line~\cite{peropadre_nonequilibrium_2013}, which is also suited for the ultrastrong coupling scheme, the presented real-space approach is preferable when a system consists of several qubits.

The presented numerical approach can simulate the dynamics with a driving field by applying a unitary gate corresponding to the qubit driving before each time-step integration.
Consequently, one can evaluate the effects of the driving to other off-resonant qubits or simulate the generation of entanglement between remote qubits connected by waveguide modes~\cite{warren_robust_2021}.
Since a wavefunction for the whole system is available in tensor-network-based simulations of the Hamiltonian dynamics, the complete information for the system is essentially accessible.
The presented approach has a potential to become one of useful approaches to evaluate and optimize pulse sequences for qubit controls.

\begin{acknowledgments}
    We thank K. Koshino for the fruitful discussions.
    The MPS simulations in this study were performed with ITensor library~\cite{fishman_itensor_2022}.
    This work was financially supported by JSPS KAKENHI, Grant No.~23K13042.
\end{acknowledgments}

\section*{Data availability}
The data that support the findings of this article are openly available at the zenodo repository \url{https://zenodo.org/records/15524291}.
The code for the controlled-bond expansion approaches are also openly available at the github repository \url{https://github.com/ShimpeiGoto/CBEAlgorithms}.

\bibliographystyle{apsrev4-2} 
%

\end{document}